\documentclass[preprint,aps]{revtex4}
\usepackage{color}
\begin{document}
\draft
\title{Gauge and Lorentz Covariant Schwinger-Dyson Equation for Fermion Propagator in Arbitrary
External Gauge Field}

\author{Xiang Chen$^1$, Ying Zhang$^1$, Fa-Min Chen$^2$, Qing Wang$^1$}

\address{$^1$Department of Physics,Tsinghua University,Beijing 100084,P.R.China\footnote{mailing address} \\
    $^2$Physics Department, University of Utah, Salt Lake City, Utah 84112, USA}

\date{April 24, 2007}

\begin{abstract}
 A formalism of gauge and Lorentz covariant  Schwinger-Dyson equation is built up
 for fermion propagator in presence of arbitrary external gauge field within ladder approximation. Different
external electromagnetic field dependent trial solutions are
investigated which spontaneously break chiral symmetry.

\bigskip
PACS number(s): 11.15.Tk, 12.20-m, 12.20.Ds, 12.38.Lg
\end{abstract}
\maketitle


\vspace{1cm}

Fermion propagator, once external gauge field sets in, may exhibit
new phenomena. In literature, people focus on whether and how
external field will trigger chiral symmetry breaking
(CSB)\cite{Leung,Aexp,Miransky}. We further show next that fermion
propagator in presence of external gauge field will provides us an
alternative way to realize the gauge
 invariance of the theory and leads to well known Ward-Takahashi identity (WTI) \cite{WTI}.

 It is well-known that fermion propagator is not explicitly gauge covariant.
 This is due to the fact that gauge transformation relates different Green's functions, just fermion propagator
 itself cannot realize gauge covariance.
Although the fermion propagator is not a gauge invariant quantity,
it satisfies constraints of gauge covariance. It obeys WTI which
relates it to the fermion-boson vertex. Moreover, it follows the
gauge covariance relation dictated by its Landau-Khalatnikov-Fradkin
transformation law \cite{LKF}. Now with fermion propagator in
presence of external gauge field,
 we present a way to realize the gauge covariance for fermion propagator itself. Suppose $S(x,y;A)$ is a
 fermion propagator in presence of external gauge field $A^\mu$ which can be either Abelian or non-abelian,
\begin{eqnarray}
S(x,y;A)=-i\langle 0|{\bf T}\psi(x)\overline{\psi}(y)|0\rangle_A\;, \label{physicalS}
\end{eqnarray}
 then $S(x,y;0)$ is usual fermion propagator without external gauge field, which is
 function of $x-y$ due to invariance of space-time translation. With its Fourior transformation
 $S(p)$,
\begin{eqnarray}
S(x,y;0)=\int\frac{d^4p}{(2\pi)^4}e^{-ip\cdot(x-y)}~S(p)=S(i\partial_x)\delta^4(x-y)\;,\label{S(p)def}
\end{eqnarray}
we can construct a gauge covariant fermion propagator $S(x,y;A)$,
\begin{eqnarray}
S(x,y;A)=S(i\nabla_x)\delta(x-y)\;,\hspace{2cm}\nabla^{\mu}_x=\partial^{\mu}_x-igA^{\mu}(x)\;,\label{Scovariantdef}
\end{eqnarray}
which is covariant under gauge transformation
$gA^{\mu}(x)\rightarrow
V(x)gA^{\mu}(x)V^{\dagger}(x)-iV(x)[\partial^{\mu}V^{\dagger}(x)]$,
or in terms of covariant derivative $\nabla^\mu_x\rightarrow
V(x)\nabla^{\mu}_xV^\dagger(x)$, it transforms as
\begin{eqnarray}
S(x,y;A)\rightarrow S'(x,y;A)=
V(x)S(i\nabla_x)V^\dagger(x)\delta^4(x-y)=V(x)S(x,y;A)V^{\dagger}(y)\label{Strans0}\;.
\end{eqnarray}
Note here what we have introduced into theory is gauge potential
$A^\mu$,
 it has a special non-homogeneous term $-(i/g)V\partial^{\mu}V^{\dagger}$ under gauge transformation which play the
 key role in covariant differential and then can make the theory covariant
under gauge transformations. This is different as those discussions
given by Leung {\em et. al.} \cite{Leung0},
 where external magnetic field is introduced into theory, since
 magnetic field as field strength of Abelian gauge theory is a gauge invariant quantity, add it into theory can not
  make theory gauge covariant.

There is ambiguity when we naively generalize from $S(p)$ to
$S(i\nabla_x)$: the different component of $p^\mu$ are commutative,
but $\nabla^\mu$ are not with
$[\nabla_x^\mu,\nabla^\nu_x]=-iF^{\mu\nu}_x$. Which cause
uncertainties in the arrangement of different sequences of
$S(i\nabla_x)$'s arguments. For example, a term of $(p^2)^2$ can
have different generalizations such as $(-\nabla^2_x)^2$ or
$\nabla^\mu_x\nabla^\nu_x\nabla_{\mu,x}\nabla_{\nu,x}$, which are
not the same due to their nonzero difference
\begin{eqnarray}
(-\nabla^2_x)^2-\nabla^\mu_x\nabla^\nu_x\nabla_{\mu,x}\nabla_{\nu,x}
=\nabla^\mu_x[\nabla_{\mu,x},\nabla_{\nu,x}]\nabla^\nu_x=-i\nabla^\mu_xF_{\mu\nu,x}\nabla^\nu_x\;.\nonumber
\end{eqnarray}
Although it is not one to one correspondence when we generalize from $S(p)$ to $S(i\nabla_x)$, the inverse process
is unique. So as long as there is a way to determine $S(i\nabla_x)$, we can uniquely fix conventional fermion
propagator $S(p)$ by just vanishing external gauge field $A^\mu$.

Once  fermion propagator with external gauge field is known, we can
obtain all Green's functions for two fermions and fermion currents
by differentiating $S(x,y;A)$ with respect to $A^\mu_\alpha$,
\begin{eqnarray}
-i\langle 0|{\bf T}\psi(x)\overline{\psi}(y) g\overline{\psi}(x_1)t_{\alpha_1}\!\gamma^{\mu_1}\!\psi(x_1)\cdots
g\overline{\psi}(x_n)t_{\alpha_n}\!\gamma^{\mu_n}\!\psi(x_n)|0\rangle = \frac{\delta S(x,y;A)}{\delta
A^{\mu_1}_{\alpha_1}\!(x_1)\cdots\delta A^{\mu_n}_{\alpha_n}\!(x_n)}\bigg|_{A^\mu_\alpha=0}\;,~~~\label{Gn}
\end{eqnarray}
where $t_\alpha$ is gauge group generator. This imply that an
infinite series of Green's functions are known through $S(x,y;A)$.
With it, original relations  build up by WT identities from gauge
invariance of the theory now are explicitly exhibited through
external gauge field dependence of the fermion propagator. To verify
our statement, as a simplified example, we apply relation
(\ref{Strans0}) to abelian gauge theory in the following to
explicitly derive WT identities.

 Consider the transformation matrix $V(x)=e^{i\beta(x)}$ in abelian gauge theory, then (\ref{Strans0}) imply
$S(x,y;A')=V(x)S(x,y;A)V^{\dagger}(y)$, which for infinitesimal $\beta(x)$ become
\begin{eqnarray}
S(x,y;A-\frac{1}{g}\partial\beta)=S(x,y;A)+i\beta(x)S(x,y;A)-iS(x,y;A)\beta(y)\label{WTI0}
\end{eqnarray}
Expand the l.h.s. of above equation as
\begin{eqnarray}
S(x,y;A-\frac{1}{g}\partial\beta)=S(x,y;A)-\frac{1}{g}\int d^4z\frac{\delta S(x,y;A)}{\delta
A^{\sigma}_{\alpha}(z)}\frac{\partial\beta_{\alpha}(z)}{\partial z_{\sigma}}\label{WTI1}
\end{eqnarray}
Ignoring total derivative term, compare order $\beta$ terms in (\ref{WTI0}) and (\ref{WTI1}), we find
\begin{eqnarray}
i[\delta(x-z)-\delta(x-y)]S(x,y;A)=\frac{1}{g}\frac{\partial}{\partial z_{\sigma}}\frac{\delta S(x,y;A)}{\delta
A^{\sigma}_{\alpha}(z)}
\end{eqnarray}
Combine with definition (\ref{physicalS}) and expression (\ref{Gn}),
above equation reduce to the standard WT identity
\begin{eqnarray}
&&\hspace{-1.2cm}[\delta(x-z)-\delta(x-y)]\langle 0|{\bf
T}\psi(x)\overline{\psi}(y)|0\rangle_A=-i\frac{\partial}{\partial z_{\sigma}} \langle 0|{\bf
T}\psi(x)\overline{\psi}(y)\overline{\psi}(z)t_{\alpha}\gamma_{\sigma}\psi(z)|0\rangle_A
\end{eqnarray}
One can obtain a series WT identities by further differentiating
with external gauge fields and then vanishing them.

To obtain fermion propagator with external gauge field, we need to
solve the corresponding dynamical equation. As we know, fermion
propagator in the field theory is determined dynamically by
Schwinger-Dyson equation (SDE), now what we need to do is to
generalize traditional SDE for fermion propagator to the case
including external gauge field and build up a covariant formalism
which can keep the gauge covariance lead by external gauge field.
This gauge covariance of external field is very important since it
is directly related the validity of WT identities. It is purpose of
this paper to build up a gauge covariant SDE and then search for its
 gauge covariant solutions. This work can be seen
as a generalization of our previous result on gauge and Lorentz
covariant computation program for free fermion propagator in
presence of external gauge field\cite{YHAn,Gitman}. For simplicity,
we only consider massless Abelian gauge theory as a prototype of the
discussion. Since bare mass of fermion vanishes, the Lagrangian of
this theory exhibit chiral symmetry. Our plan is first to show that,
in contrast to conventional case that ladder approximation SDE is
not gauge invariant, include in external electromagnetic potential
$A_{\mu}$ will really make it gauge invariant under gauge
transformation of $A_{\mu}$ and we construct systematic method to
solve this equation gauge and Lorentz covariantly. Then we will try
to search for possible CSB solutions. Since this paper mainly focus
on the issue of constructing gauge and Lorentz covariant SDE for
fermion propagator in presence of arbitrary external gauge field,
for the CSB solution we limit ourself in emphasizing that our
formalism could include various possible solutions and do not make
further detail discussions.

 \vspace*{0.5cm}We start by considering following standard ladder approximation SDE for fermion
propagator in massless QED in Minkovski space,
\begin{eqnarray}
&&S^{-1}(x,y;0)-i\partial\!\!\!/_x\delta^4(x-y)=g^2D^{\mu\nu}(x,y)\gamma_{\mu}S(x,y;0)\gamma_{\nu}\;,\label{SDE0}
\end{eqnarray}
with photon propagator $D^{\mu\nu}(x,y)=\int\frac{d^4p}{(2\pi)^4}e^{-ip\cdot(x-y)}D^{\mu\nu}(p)$. With
(\ref{S(p)def}), SDE (\ref{SDE0}) can be written in terms of momentum space fermion propagator $S(p)$,
\begin{eqnarray}
[S^{-1}(i\partial_x)-i\partial\!\!\!/_x]\delta^4(x-y)=g^2D^{\mu\nu}(x,y)\gamma_{\mu}S(i\partial_x)\delta(x-y)\;.
\label{SDE1}
\end{eqnarray}
Now we generalize SDE (\ref{SDE0}) to the case in presence of external electromagnetic field by replacing
$\partial\!\!\!/_x$ in l.h.s. of (\ref{SDE0}) with covariant derivative $\nabla\!\!\!\!/\;_x$,
\begin{eqnarray}
S^{-1}(x,y;A)-i\nabla\!\!\!\!/\;_x\delta^4(x-y)=g^2D^{\mu\nu}(x,y)\gamma_{\mu}S(x,y;A)\gamma_{\nu}\;,\label{SDE2}
\end{eqnarray}
and we are looking for its type (\ref{Scovariantdef}) solution. Before constructing covariant computation
formalism, we first show that this equation is invariant under gauge transformation (\ref{Strans0}). This can be
done with help of transformation law for $S^{-1}(x,y;A)\rightarrow
S^{-1\prime}(x,y;A)=V(x)S^{-1\prime}(x,y;A)V^{\dagger}(y)$, since it will result correct relation
\begin{eqnarray}
&&\hspace{-0.5cm}\int d^4y~S^{-1\prime}(x,y;A)S'(y,z;A)=\int
d^4y~V(x)S^{-1}(x,y;A)S(y,z;A)V^{\dagger}(z)=\delta^4(x-z)\;.~~~~
\end{eqnarray}
Then (\ref{SDE2}) can be rewritten by multiply $V(x)$ from l.h.s. and $V^\dagger(y)$ from r.h.s. of the equation
\begin{eqnarray}
V(x)S^{-1}(x,y;A)V^\dagger(y)-iV(x)\nabla\!\!\!\!/\;_x\delta^4(x-y)V^\dagger(y)
=g^2D^{\mu\nu}(x,y)\gamma_{\mu}V(x)S(x,y;A)V^\dagger(y)\gamma_{\nu}\;.~~~~\label{SDE3}
\end{eqnarray}
Since the second term of l.h.s. of (\ref{SDE3}) is equal to
$-iV(x)\nabla\!\!\!\!/\;_x\delta^4(x-y)V^\dagger(x)=-i\nabla\!\!\!\!/\;'_x\delta^4(x-y)$, it is easy to see
(\ref{SDE3}) is just SDE for fermion propagator after gauge transformation
\begin{eqnarray}
S^{-1\prime}(x,y;A)-i\nabla_x'\delta^4(x-y)=g^2D^{\mu\nu}(x,y)\gamma_{\mu}S'(x,y;A)\gamma_{\nu}\;.\label{SDE4}
\end{eqnarray}
Which achieve the gauge invariance of SDE (\ref{SDE2}).

To obtain gauge covariant solution of (\ref{SDE2}), note that (\ref{Scovariantdef}) implies that $S(x,y;A)$
expressed in terms of $S(i\nabla_x)$ can keep the gauge covariance, so we can use (\ref{SDE2}) to fix the form of
$S(i\nabla_x)$. To realize this idea, we need first know how to express $S^{-1}(x,y;A)$ appeared in (\ref{SDE2})
in terms of $S(i\nabla_x)$. We make our attempt by introducing $S^{-1}(i\nabla_x)$ similar as
(\ref{Scovariantdef})
\begin{eqnarray}
S^{-1}(x,y;A)=S^{-1}(i\nabla_x)\delta(x-y)\;,\label{S-1covariantdef}
\end{eqnarray}
which has an alternative expression
\begin{eqnarray}
S^{-1}(x,y;A)=\!\int\!\frac{d^4q}{(2\pi)^4}e^{-iq\cdot(x-y)}S^{-1}(i\nabla_x\!\!+\!q)
=\!\int\!\frac{d^4q}{(2\pi)^4}e^{-iq\cdot(x-y)}S^{-1}(i\nabla_y\!\!+\!q\!-\!i\partial_y)\bigg|_{i\partial_y\rm
moved\;to\;l.h.s}\;,~~\label{S-1exp00}
\end{eqnarray}
where we have used property that
\begin{eqnarray}
\int\frac{d^4q}{(2\pi)^4}e^{-iq\cdot(x-y)}f(q)g(x)&=&f(-i\partial_y)\int\frac{d^4q}{(2\pi)^4}e^{-iq\cdot(x-y)}g(x)
=f(-i\partial_y)[g(y)\delta^4(x-y)]\nonumber\\
&=&\int\frac{d^4q}{(2\pi)^4}e^{-iq\cdot(x-y)}f(q-i\partial_y)g(y)\;.\nonumber
\end{eqnarray}
With (\ref{S-1exp00}), we can further express $S^{-1}(x,y;A)$ as
\begin{eqnarray}
S^{-1}(x,y;A)=\int\frac{d^4q}{(2\pi)^4}S^{-1}(-i\overleftarrow{\nabla}_y+q)e^{-iq\cdot(x-y)}=\delta(x-y)S^{-1}
(-i\overleftarrow{\nabla}_y)\;,\label{S-1exp0}
\end{eqnarray}
where $\overleftarrow{\nabla}^\mu_x=\overleftarrow{\partial}^\mu_x+igA^\mu(x)$. With help of expression
(\ref{S-1exp0}),
\begin{eqnarray}
\delta(x-y)&=&\int d^4x'~S(x,x')S^{-1}(x',y)
=\int d^4x'S(i\nabla_x)\delta(x-x')\delta(x'-y)S^{-1}(-i\overleftarrow{\nabla}_y)\nonumber\\
&=&S(i\nabla_x)\delta(x-y)S^{-1}(-i\overleftarrow{\nabla}_y) =S(i\nabla_x)S^{-1}(i\nabla_x)\delta(x-y)\;,~~~~
\end{eqnarray}
which imply the constraint for $S^{-1}(i\nabla_x)$ is
\begin{eqnarray}
S(i\nabla_x)S^{-1}(i\nabla_x)=1\;.\label{S-1Sidentity}
\end{eqnarray}
With this type of $S^{-1}(i\nabla_x)$ and (\ref{Scovariantdef}), SDE (\ref{SDE2}) can be expressed as a equation
to fix $S(i\nabla_x)$
\begin{eqnarray}
[S^{-1}(i\nabla_x)-i\nabla\!\!\!\!/\;_x]\delta^4(x-y)=g^2D^{\mu\nu}(x,y)\gamma_{\mu}
S(i\nabla_x)\delta^4(x-y)\gamma_{\nu}\;,\label{SDE5}
\end{eqnarray}
it goes back to SDE (\ref{SDE1}) when we switching off external
electromagnetic potential $A^\mu$.

To solve SDE (\ref{SDE5}), a direct method is to expand the solution
in terms of powers of external electromagnetic potential $A^\mu$ and
take special Schwinger-Fock gauge to relate gauge potential
$A^\mu(x)$ directly with gauge field strength $F^{\mu\nu}_x$
\cite{Aexp}. This expansion is not gauge covariant, since it
decompose covariant derivative apart and the operation for various
parts of the covariant derivative are different which will violet
the gauge covariance of the computation program. So this method will
generally result a non-covariant $S(x,y;A)$ due to the non-covariant
computation procedure. To keep gauge covariance of $S(x,y;A)$, we
need to construct a covariant solution formalism. Inspired by the
method proposed in Ref.\cite{YHAn}, we write $S(i\nabla_x)$ as
\begin{eqnarray}
S(x,y;A)&=&S(i\nabla_x)\delta(x-y)=\int\frac{d^4p}{(2\pi)^4}e^{-ip\cdot(x-y)} S(i\nabla_x+p)1\nonumber\\
&&\hspace{-2.5cm}=\int\frac{d^4p}{(2\pi)^4}e^{-ip\cdot(x-y)}e^{i\nabla_x\cdot\frac{\partial}{\partial p}}
S(p-\tilde{F}_{p,x}) e^{-i\nabla_x\cdot\frac{\partial}{\partial p}}1
=\int\frac{d^4p}{(2\pi)^4}e^{-ip\cdot(x-y)}e^{i\nabla_x\cdot\frac{\partial}{\partial p}}
S(p-\tilde{F}_{p,x})1\;.~~~~~\label{Sexp0}
\end{eqnarray}
$S^{-1}(i\nabla_x)$ has the similar result. $\tilde{F}^\mu_{p,x}$ is defined as
\begin{eqnarray}
\tilde{F}^\mu_{p,x}&=&p^\mu-e^{-i\nabla_x\cdot\frac{\partial}{\partial
p}}(p^\mu+i\nabla^\mu_x)e^{i\nabla_x\cdot\frac{\partial}{\partial p}}\nonumber\\
&=&-\frac{1}{2}[\nabla^\nu_x,\nabla^\mu_x]\frac{\partial}{\partial
p^\nu}+\frac{i}{3}\partial^{\lambda}[\nabla^\nu_x,\nabla^{\mu}_x]\frac{\partial^2}{\partial p^{\lambda}\partial
p^{\nu}}+\frac{1}{8}\partial^{\mu'}\partial^{\nu'}[\nabla^\nu_x,\nabla^{\mu}_x]\frac{\partial^3}{\partial
p^{\mu'}\partial p^{\nu'}\partial p^{\nu}}+O(p^5)\;,~~~~~\label{tildeF}
\end{eqnarray}
in which for convenience of the calculation we assign powers to $\nabla^\mu$ and it is taken as order of $p^1$ and
then $O(p^5)$ denote terms with at least five covariant derivatives. Note $\tilde{F}^\mu_{p,x}$ commute with
$\tilde{F}^\nu_{p,x}$, but this property no longer holds when we goes to nonabelian gauge theory. Further
\begin{eqnarray}
[p^\mu-\tilde{F}^\mu_{p,x},p^\nu-\tilde{F}^\nu_{p,x}]=e^{-i\nabla_x\cdot\frac{\partial}{\partial
p}}[p^\mu+i\nabla^\mu_x,p^\nu+i\nabla^\nu_x]e^{i\nabla_x\cdot\frac{\partial}{\partial p}} \stackrel{\mbox{\tiny
abelian}}{====}iF^{\mu\nu}_x\;.\label{commutator}
\end{eqnarray}
The l.h.s. of (\ref{SDE5}) now become
\begin{eqnarray}
&&\hspace{-0.5cm}[S^{-1}(i\nabla_x)-i\nabla\!\!\!\!/\;_x]\delta^4(x-y)=\int\frac{d^4p}{(2\pi)^4}e^{-ip\cdot
(x-y)}\bigg[e^{i\nabla_x\cdot\frac{\partial}{\partial p}}S^{-1}(p-\tilde{F}_{p,x})1-
i\nabla\!\!\!\!/\;_x-p\!\!\!/\;\bigg]\nonumber\\
&&\hspace{-0.5cm}=\int\frac{d^4p}{(2\pi)^4}e^{-ip\cdot(x-y)} e^{i\nabla_x\cdot\frac{\partial}{\partial
p}}[S^{-1}(p-\tilde{F}_{p,x})1-p\!\!\!/\;]\;.\nonumber
\end{eqnarray}
While r.h.s. of (\ref{SDE5}) become
\begin{eqnarray}
&&\hspace{-0.6cm}g^2D^{\mu\nu}(x,y)\gamma_{\mu}S(i\nabla_x)\delta^4(x-y)\gamma_{\nu}\nonumber\\
&&\hspace{-0.6cm}=g^2\int\frac{d^4pd^4q}{(2\pi)^8}e^{-i(q+p)\cdot(x-y)} D^{\mu\nu}(q)
\bigg[e^{i\nabla_x\cdot\frac{\partial}{\partial p}}\gamma_{\mu}S(p-\tilde{F}_{p,x})1\gamma_{\nu}\bigg]\nonumber\\
&&\hspace{-0.6cm}=g^2\int\frac{d^4p}{(2\pi)^4}e^{-ip\cdot(x-y)}\bigg[e^{i\nabla_x\cdot\frac{\partial}{\partial
p}}\int\frac{d^4q}{(2\pi)^4} D^{\mu\nu}(q)\gamma_{\mu}S(p-q-\tilde{F}_{p-q,x})1\;\gamma_\nu\bigg]\;.\nonumber
\end{eqnarray}
Combine l.h.s. and r.h.s. of (\ref{SDE5}) results together, subtract out factor
$\int\frac{d^4p}{(2\pi)^4}e^{-ip\cdot(x-y)}~e^{i\nabla_x\cdot\frac{\partial}{\partial p}}$ in both sides of the
equation, we obtain
\begin{eqnarray}
S^{-1}(p-\tilde{F}_{p,x})1-p\!\!\!/\;=g^2\int\frac{d^4q}{(2\pi)^4}
D^{\mu\nu}(p-q)\gamma_{\mu}S(q-\tilde{F}_{q,x})1\;\gamma_\nu\;.\label{SDE6}
\end{eqnarray}
The advantage of above version of SDE in presence of external electromagnetic field is that external field
appeared in the equation is through field strength not original gauge potential, it is this feature of the
formalism making our computation program gauge covariant. To solve above equation, we notice that
(\ref{S-1Sidentity}) now imply constraint
\begin{eqnarray}
S(p-\tilde{F}_{p,x})S^{-1}(p-\tilde{F}_{p,x})=1\;.\label{S-1Sidentity1}
\end{eqnarray}
with it, in the following we try to search for its solutions.
\subsection{Trivial solution}

it is easy to see that conventional fermion propagator which satisfy
traditional SDE
\begin{eqnarray}
[Z(p^2)-1]p\!\!\!
/\;-\Sigma(p^2)=g^2\int\frac{d^4q}{(2\pi)^4}D^{\mu\nu}(p-q)\gamma_{\mu}\frac{1}{Z(q^2)q\!\!\!
/\;-\Sigma(q^2)}\gamma_\nu\;,\label{SDE7}
\end{eqnarray}
just coincide (\ref{SDE6}) and (\ref{S-1Sidentity1}) if we identify
\begin{eqnarray}
S^{-1}(p-\tilde{F}_{p,x})1=Z(p^2)p\!\!\! /\;-\Sigma(p^2)\hspace{2cm}
S(p-\tilde{F}_{p,x})1=\frac{1}{Z(p^2)p\!\!\! /\;-\Sigma(p^2)}\;,
\label{S-1Ssolution}
\end{eqnarray}
which seems strange at first sight, since l.h.s. of
(\ref{S-1Ssolution}) depend on argument $p^\mu-\tilde{F}^\mu$ which
require r.h.s of (\ref{S-1Ssolution}) should be written as some
function of argument $p^\mu-\tilde{F}^\mu$, or one can finish
operations of momentum differentials inside $\tilde{F}^\mu$. Since
from (\ref{tildeF}), these differentials has coefficients depend on
external field strength, finishing operations of momentum
differentials in general will leave some momentum and external field
strength dependent terms, but now r.h.s. of (\ref{S-1Ssolution})
only depend on momentum, external field strength donot show up.
There must be some special reason. Notice that since
$[p^\mu,\tilde{F}^\nu_{p,x}]\neq 0$, as we discussed before for
$S(i\nabla_x)$, in general there will be arrangement ambiguity for
different sequences of $S^{-1}(p-\tilde{F})$'s and
$S(p-\tilde{F})$'s argument $p^\mu-\tilde{F}^\mu$ . Now r.h.s. of
(\ref{S-1Ssolution}) in fact offers a definition of the operator
orders. We show next that carefully arrange the order of argument of
$p^\mu-\tilde{F}^\mu$, it is possible to realize above strange
result. As an example, if we perform momentum expansion, for $p^2$
term, there is no arrangement order problem
 \begin{eqnarray}
(p\!-\!\tilde{F}_{p,x})^21=\bigg[p^2+\frac{1}{2}[\nabla^\nu_x,\nabla^\mu_x](\frac{\partial}{\partial
p^\nu}p_\mu\! +\!p_\mu\frac{\partial}{\partial
p^\nu})+\frac{1}{4}[\nabla^\nu_x,\nabla^\mu_x][\nabla^{\nu'}_x,\nabla_{\mu,x}]
\frac{\partial^2}{\partial p^\nu\partial
p^{\nu'}}\bigg]1=p^2\;.\nonumber
\end{eqnarray}
While for $p^4$ term we have three different kinds of arrangements,
the first is
\begin{eqnarray}
&&\hspace{-0.6cm}(p\!-\!\tilde{F}_{p,x})^2(p\!-\!\tilde{F}_{p,x})^21=\bigg[p^2+\frac{1}{2}[\nabla^\nu_x,\nabla^\mu_x](\frac{\partial}{\partial
p^\nu}p_\mu\!+\!p_\mu\frac{\partial}{\partial
p^\nu})+\frac{1}{4}[\nabla^\nu_x,\nabla^\mu_x][\nabla^{\nu'}_x,\nabla_{\mu,x}]
\frac{\partial^2}{\partial p^\nu\partial
p^{\nu'}}\bigg]^21\nonumber\\
&&\hspace{3.2cm}=p^4+\frac{1}{2}[\nabla^\nu_x,\nabla^\mu_x][\nabla_{\nu,x},\nabla_{\mu,x}]\;,\label{arrange1}
\end{eqnarray}
the second is
\begin{eqnarray}
&&\hspace{-0.6cm}(p^\mu\!-\!\tilde{F}_{p,x}^\mu)(p^\nu\!-\!\tilde{F}_{p,x}^\nu)
(p_\mu\!-\!\tilde{F}_{\mu,p,x})(p_\nu\!-\!\tilde{F}_{\nu,p,x})1\nonumber\\
&&\hspace{-0.6cm}=(p\!-\!\tilde{F}_{p,x})^2(p\!-\!\tilde{F}_{p,x})^21+
(p^\mu\!-\!\tilde{F}_{p,x}^\mu)[(p^\nu\!-\!\tilde{F}_{p,x}^\nu),
(p_\mu\!-\!\tilde{F}_{\mu,p,x})](p_\nu\!-\!\tilde{F}_{\nu,p,x})1\nonumber\\
&&\hspace{-0.6cm}=(p\!-\!\tilde{F}_{p,x})^2(p\!-\!\tilde{F}_{p,x})^21
-\frac{1}{2}[\nabla^\nu_x,\nabla^\mu_x][\nabla_{\nu,x},\nabla_{\mu,x}]=p^4\;,
\end{eqnarray}
where we have used result (\ref{commutator}). The third is
\begin{eqnarray}
&&\hspace{-0.6cm}(p^\mu\!-\!\tilde{F}_{p,x}^\mu)(p\!-\!\tilde{F}_{p,x})^2
(p_\mu\!-\!\tilde{F}_{\mu,p,x})1\nonumber\\
&&\hspace{-0.6cm}=(p^\mu\!-\!\tilde{F}_{p,x}^\mu)(p^\nu\!-\!\tilde{F}_{p,x}^\nu)
(p_\mu\!-\!\tilde{F}_{\mu,p,x})(p_\nu\!-\!\tilde{F}_{\nu,p,x})1
-(p^\mu\!-\!\tilde{F}_{p,x}^\mu)(p^\nu\!-\!\tilde{F}_{p,x}^\nu)
[(p_\mu\!-\!\tilde{F}_{\mu,p,x}),(p_\nu\!-\!\tilde{F}_{\nu,p,x})]1\nonumber\\
&&\hspace{-0.6cm}=p^4-\frac{1}{2}[\nabla^\nu_x,\nabla^\mu_x][\nabla_{\nu,x},\nabla_{\mu,x}]\;.
\end{eqnarray}
Therefore there are two ways of arrangement matching our result
(\ref{S-1Ssolution}), i.e. make field strength disappear, we can
adjust its relative ratio to cancel the external field dependence
\begin{eqnarray}
&&\bigg[(\frac{1}{2}-a)[(p\!-\!\tilde{F}_{p,x})^2(p\!-\!\tilde{F}_{p,x})^2+(p^\mu\!-\!\tilde{F}_{p,x}^\mu)(p\!-\!\tilde{F}_{p,x})^2
(p_\mu\!-\!\tilde{F}_{\mu,p,x})]\nonumber\\
&&+a(p^\mu\!-\!\tilde{F}_{p,x}^\mu)(p^\nu\!-\!\tilde{F}_{p,x}^\nu)
(p_\mu\!-\!\tilde{F}_{\mu,p,x})(p_\nu\!-\!\tilde{F}_{\nu,p,x})\bigg]1=p^4\;,~~~~~~\label{p4arrange}
\end{eqnarray}
in which parameter $a$ is an arbitrary constant which can be tuned
to make both $p^4$ terms appeared in $S(p-\tilde{F}_{p,x})1$ and
$S^{-1}(p-\tilde{F}_{p,x})1$ be all independent on external field.
One can continue to discuss all high orders which make us believe
the correctness of solution (\ref{S-1Ssolution}). Now we already
explain that why (\ref{S-1Ssolution}) is the solution of SDE with
presence of external field. This solution at level of
$S(p-\tilde{F}_{p,x})1$ and $S^{-1}(p-\tilde{F}_{p,x})1$ is
independent of external field.

With result (\ref{S-1Ssolution}) and similar derivation as
(\ref{Sexp0}), we obtain explicit expression for $S(x,y;A)$ in terms
of commutators as
\begin{eqnarray}
&&\hspace{-0.5cm}
S^{-1}(x,y;A)=\int\frac{d^4p}{(2\pi)^4}e^{-ip\cdot(x-y)}e^{i\nabla_x\cdot\frac{\partial}{\partial
p}}
[Z(p^2)p\!\!\!/\;-\Sigma(p^2)]\nonumber\\
&&\hspace{-0.5cm}S(x,y;A)=\int\frac{d^4p}{(2\pi)^4}e^{-ip\cdot(x-y)}e^{i\nabla_x\cdot\frac{\partial}{\partial
p}} \frac{1}{Z(p^2)p\!\!\!/\;-\Sigma(p^2)}\;.\label{Sfinal}
\end{eqnarray}
Exploit the function ${\rm\bf
a}[x-y;A(x)]=e^{-z\cdot\nabla_x}1\bigg|_{z=x-y}$ introduced in
Ref.\cite{YHAn}, using partial differential rules changing the
object of differential $\frac{\partial}{\partial p}$ from
$[Z(p^2)p\!\!\!/\;-\Sigma(p^2)]^{-1}$ to $e^{-ip\cdot(x-y)}$ and
note that $e^{-i\nabla_x\cdot\frac{\partial}{\partial
p}}e^{-ip\cdot(x-y)}={\rm\bf a}[x-y;A(x)]e^{-ip\cdot(x-y)}$, we can
rewrite (\ref{Sfinal}) as
\begin{eqnarray}
S(x,y;A)={\rm\bf a}[x-y;A(x)]S(x,y;0)+\mbox{momentum space total
derivative terms}\;,
\end{eqnarray}
where phrase momentum space total derivative terms appeared in above
formula denote terms like
\begin{eqnarray}
\int\frac{d^4p}{(2\pi)^4}\frac{\partial}{\partial
p_\mu}\bigg[e^{-ip\cdot(x-y)}f[-z\cdot\nabla_x,\nabla_x\cdot\frac{\partial}{\partial
p}] \nabla^\mu_x
g[-z\cdot\nabla_x,\nabla_x\cdot\frac{\partial}{\partial p}]
[Z(p^2)p\!\!\!/\;-\Sigma(p^2)]^{-1}\bigg]\;,\nonumber
\end{eqnarray}
with $f(k,l)$ and $g(k,l)$ are $k^nl^m$ type power functions. For
$x\neq y$, it is expected that the integration over total derivative
term will equivalent to the integration on the boundary of the
space-time for which the $e^{-ip\cdot(x-y)}$ in will cause strong
oscillation interference cancellation. While for the coincidence
limit of the propagator, with definition (\ref{physicalS}) and
result (\ref{Sfinal}), we find
\begin{eqnarray}
\langle 0|{\bf
T}\psi(x)\overline{\psi}(x)|0\rangle_A=-\frac{1}{4}\langle\overline{\psi}\psi\rangle
1+
i\int\frac{d^4p}{(2\pi)^4}[e^{i\nabla_x\cdot\frac{\partial}{\partial
p}}-1]\frac{1}{Z(p^2)p\!\!\!/\;-\Sigma(p^2)}\;,
\end{eqnarray}
in which the first term $\langle\overline{\psi}\psi\rangle$ is the
standard fermion condensate constant which is external
electromagnetic field and space-time coordinates independent,  the
second term is the momentum space total derivative terms. Taking
trace for spinor index, we find the difference of fermion condensate
in presence of external electromagnetic field
$\langle\overline{\psi}\psi\rangle_A$ with conventional fermion
condensate $\langle\overline{\psi}\psi\rangle$ is
\begin{eqnarray}
\langle\overline{\psi}\psi\rangle_A-\langle\overline{\psi}\psi\rangle=
-4i\int\frac{d^4p}{(2\pi)^4}~[\cosh(i\nabla_x\!\cdot\!\frac{\partial}{\partial
p})-1]~\frac{\Sigma(p^2)}{Z^2(p^2)p^2-\Sigma^2(p^2)}\;.\label{trivialresult}
\end{eqnarray}
If we further ignore the total differential terms in the integrand
of (\ref{trivialresult}), r.h.s. of above equation vanishes. Then
external electromagnetic field will play no role in the fermion
condensate, this is the reason that why we call this solution the
trivial solution.
\subsection{Solution with weak external electromagnetic field }

Beyond trivial solution discussed above, there are many nontrivial
solutions which in contrast to trivial solution, will depend on
external electromagnetic field at level of $S(p-\tilde{F}_{p,x})1$
and $S^{-1}(p-\tilde{F}_{p,x})1$. We focus our attention on the
solution for which external electromagnetic field is weak. We take
following ansatz for $S^{-1}(p-\tilde{F}_{p,x})$ and
$S(p-\tilde{F}_{p,x})$
\begin{eqnarray}
S^{-1}(p-\tilde{F}_{p,x})=p\!\!\!
/-\tilde{\Sigma}(p^2,F^2_x)\hspace{2cm}
S(p-\tilde{F}_{p,x})1=\frac{1}{p\!\!\!
/\;-\tilde{\Sigma}(p^2,F^2_x)}\;. ~~~~~\label{S-1exp1}
\end{eqnarray}
With definition
$F^2_x\equiv-[\nabla_{\nu,x},\nabla_{\mu,x}][\nabla^\nu_x,\nabla^\mu_x]$.
 Compare it to (\ref{S-1Ssolution}) in last subsection, $Z(p^2)$ is replaced to 1 which is just
for simplicity and fermion self energy $\Sigma(p^2)$ is replaced to
external electromagnetic field strength dependent
$\tilde{\Sigma}(p^2,F^2_x)$. This is due to fact that the most
general arrangement on the argument of $p-\tilde{F}_{p,x}$ in
$S^{-1}(p-\tilde{F}_{p,x})1$ and $S(p-\tilde{F}_{p,x})1$ should be
dependent on $F^2_x$. For example, the most general $p^4$ order
arrangement of the argument is not (\ref{p4arrange}), but
\begin{eqnarray}
&&\bigg[(1-a-b)(p\!-\!\tilde{F}_{p,x})^2(p\!-\!\tilde{F}_{p,x})^2+b(p^\mu\!-\!\tilde{F}_{p,x}^\mu)(p\!-\!\tilde{F}_{p,x})^2
(p_\mu\!-\!\tilde{F}_{\mu,p,x})]\nonumber\\
&&+a(p^\mu\!-\!\tilde{F}_{p,x}^\mu)(p^\nu\!-\!\tilde{F}_{p,x}^\nu)
(p_\mu\!-\!\tilde{F}_{\mu,p,x})(p_\nu\!-\!\tilde{F}_{\nu,p,x})\bigg]1=p^4+\frac{1}{2}(1-a-2b)F^2_x\;,~~~~~~\label{p4arrange1}
\end{eqnarray}
which is dependent on external field with arbitrary coefficients $a$
and $b$.  With ansatz (\ref{S-1exp1}), (\ref{SDE6}) now become
\begin{eqnarray}
-\Sigma(p^2,F^2_x)= g^2\int\frac{d^4q}{(2\pi)^4}
D^{\mu\nu}(p-q)\gamma_{\mu}\frac{1}{q\!\!\!
/\;-\Sigma(q^2,F^2_x)}\gamma_{\nu}\;,\label{SDEweak}
\end{eqnarray}
which can be decomposed into two equations
\begin{eqnarray}
&&\hspace{-1cm}0=g^2\int\frac{d^4q}{(2\pi)^4}
D^{\mu\nu}(p-q)\gamma_{\mu}\frac{q\!\!\!
/\;}{q^2-\Sigma^2(q^2,F^2_x)}\gamma_{\nu}\label{wave}\\
&&\hspace{-1cm}-\Sigma(p^2,F^2_x)=g^2\int\frac{d^4q}{(2\pi)^4}
D^{\mu}_{~\mu}(p-q)\frac{\Sigma(q^2,F^2_x)}{q^2
-\Sigma^2(q^2,F^2_x)}\;.\label{SigmaF2}
\end{eqnarray}
Equation (\ref{wave}) can be satisfied by suitable choice of gauge
\cite{Munczek}. (\ref{SigmaF2}) is same as traditional SDE for
fermion self energy, except the feature that it depends on external
electromagnetic field strength $F_x^2$. With this property,
conventional result for SDE can also be applied here, i.e., there
exist a critical coupling, nonzero solution happens if the coupling
is stronger than its critical value. Since we are interested in the
case of weak external field, we can further expand
$\Sigma(p^2,F^2_x)$ as
\begin{eqnarray}
\Sigma(p^2,F^2_x)=\Sigma_1(p^2)F^2_x\;,
\end{eqnarray}
weak field assumption make it legal of dropping out high order
external field terms and external field independent term in the
expansion is ignored, (\ref{SigmaF2}) in this case become
\begin{eqnarray}
 -\Sigma_1(p^2)=g^2\int\frac{d^4q}{(2\pi)^4} D^{\mu}_{~\mu}(p-q)\frac{\Sigma_1(q^2)}{q^2}\;,\label{Sigma1}
\end{eqnarray}
which is just linearized SDE for fermion self energy , then we get
solution
\begin{eqnarray}
\Sigma_1(p^2)\propto\Sigma(p^2)\bigg|_{\tiny\rm linearized}\;.
\end{eqnarray}
With it the fermion condensate now become
\begin{eqnarray}
\langle 0|{\bf
T}\psi(x)\overline{\psi}(x)|0\rangle_A=i\int\frac{d^4p}{(2\pi)^4}\frac{\Sigma_1(p^2)}{p^2}F_x^2\;.
\end{eqnarray}
We find that the fermion condensate is nonzero and linear in $F_x^2$
at weak external field situation. This result is similar as that
obtained in Ref.\cite{weak}. The feature of this kind solution is
that it has same critical coupling as that in conventional case
without external field \cite{Aexp} and condensate is proportional to
$F_x^2$ which imply that external field, when it is weak, tends to
enhance the value of the condensate.
\subsection{Possibility of finding solution with strong external electromagnetic field}

The solution discussed in last subsection has feature that it will
not generate CSB when coupling of the theory below its critical
value. In literature, people are interested in the case that CSB
happens at any couplings as long as the external electromagnetic
field is strong enough \cite{Leung,Miransky}. Now we investigate
this type solution by taking following ansatz
\begin{eqnarray}
S^{-1}(p-\tilde{F}_{p,x})1&=&[1+C_{\mu\nu}(p,F_x)\frac{\partial^2}{\partial
p_\mu\partial p_{\nu}}][p\!\!\!
/\;-\tilde{\Sigma}(p^2,F_x^2)]\label{S-1SsolutionS}\\
&=&p\!\!\!
/\;-\tilde{\Sigma}(p^2,F_x^2)-2C^{\mu}_{~\mu}(p,F_x)\tilde{\Sigma}'(p^2,F_x^2)-4C_{\mu\nu}(p,F_x)p^{\mu}p^{\nu}\tilde{\Sigma}''(p^2,F_x^2)\;,
\nonumber
\end{eqnarray}
where prime on the $\tilde{\Sigma}$ denote the differential with
respect to variable $p^2$. The difference of above ansatz with
trivial solution (\ref{S-1Ssolution}) and weak external
electromagnetic field solution (\ref{S-1exp1}) is that now we
include in the effect of momentum differentials which in trivial
solution is forced to be cancelled and in weak external field
solution is represented through its $F_x^2$ dependence by arranging
orders of the arguments for $S^{-1}$. The ansatz we take here is
just as a prototype to explain the effects from momentum
differentials, a relatively simple differential term is taken in
(\ref{S-1SsolutionS}) and we donot discuss how to realize it through
arranging the order of the argument $p^\mu-\tilde{F}_{p,x}^\mu$ as
which we have done in last two subsections.

We parameterize $C_{\mu\nu}(p,F_x)$ as
$C_{\mu\nu}(p,F_x)=F^{\mu'}_{\mu,x}F_{\mu'\nu,x}D(p,F_x)$ and define
\begin{eqnarray}
\tilde{\Sigma}(p^2,F_x^2)+2C^{\mu}_{~\mu}(p,F_x)\tilde{\Sigma}'(p^2,F_x^2)
+4C_{\mu\nu}(p,F_x)p^{\mu}p^{\nu}\tilde{\Sigma}''(p^2,F_x^2)=
e^{-C(p,F_x)}\tilde{\Sigma}(p^2,F_x^2)\;,
\end{eqnarray}
which leads
\begin{eqnarray}
D(p,F_x)=\frac{[e^{-C(p,F_x)}-1]\tilde{\Sigma}(p^2,F_x^2)}{2F^2\tilde{\Sigma}'(p^2,F_x^2)
+4p^{\mu}p^{\nu}F_{\mu'\mu,x}F^{\mu'}_{~\nu,x}\tilde{\Sigma}''(p^2,F_x^2)}\;.
\end{eqnarray}
With (\ref{S-1SsolutionS})
\begin{eqnarray}
 &&S(p-\tilde{F}_{p,x})1=\frac{1}{p\!\!\!
/\;-\tilde{\Sigma}(p^2,F_x^2)}[1+C_{\mu\nu}(p,F_x)\frac{\partial^2}{\partial
p_\mu\partial p_{\nu}}]^{-1}1 =\frac{1}{p\!\!\!
/\;-\tilde{\Sigma}(p^2,F_x^2)}\;.
\end{eqnarray}
(\ref{SDE6}) then become
\begin{eqnarray}
-\tilde{\Sigma}(p^2,F_x^2)=g^2e^{C(p,F_x)}\int\frac{d^4q}{(2\pi)^4}
D^{\mu\nu}(p-q)\gamma_{\mu}\frac{1}{p\!\!\!
/\;-\tilde{\Sigma}(p^2,F_x^2)}\gamma_\nu\;,\label{SDE8}
\end{eqnarray}
which is just SDE (\ref{SDEweak}) discussed in last subsection with
coupling constant $g^2$ being replaced with $g^2e^{C(p,F_x)}$. As
long as factor $e^{C(p,F_x)}$ become large when external field is
strong, effective coupling constant $g^2e^{C(p,F_x)}$ will become
big enough larger than its critical value which will then trigger
CSB. This offers the possibility that strong external field trigging
CSB.

 In conclusion, we have built up general gauge invariant SDE for fermion propagator after introducing
external gauge field into traditional SDE and set up systematic
method to solve this equation gauge and Lorentz covariantly in
abelian gauge theory and with ladder approximation.  With help of
this method, we find trivial solution and weak external
electromagnetic field solution which have the same critical coupling
as case of conventional CSB in absence of external electromagnetic
field. We point out possibility to construct strong external
electromagnetic field solution which will trigger CSB when external
electromagnetic field is strong enough.


\section*{Acknowledgments}

This work was  supported by National  Science Foundation of China
(NSFC) under Grant No. 10435040 and Specialized Research Fund for
the Doctoral Program of High Education of China.



\end{document}